# Modelling of dislocations in a CDW junction: interference of the CDW and the normal carriers

Alvaro Rojo Bravo [a], Tianyou Yi [b], Natasha Kirova[a,c,d,*], Serguei Brazovskii[a,d]

[a] CNRS, LPTMS, URM 8502, Univeristé Paris-sud, Orsay, 91405, France
[b] South University of Science and Technology of China, Shenzhen, Guangdong 518055, China
[c] CNRS, LPS, URM 8626, Univeristé Paris-sud, Orsay, 91405, France
[d] International Institute of Physics, 59078-400 Natal, Rio Grande do Norte, Brazil

**Abstract**

We derive and study equations for dissipative transient processes in a constraint incommensurate charge density wave (CDW) with remnant pockets or a thermal population of normal carriers. The attention was paid to give the correct conservation of condensed and normal electrons, which was problematic at presence of moving dislocation cores if working within an intuitive Ginzburg-Landau like model. We performed a numeric modelling for stationary and transient states in a rectangular geometry when the voltage $V$ or the normal current are applied across the conducting chains. We observe creation of an array of electronic vortices, the dislocations, at or close to the junction surface; their number increases stepwise with increasing $V$. The dislocation core strongly concentrates the normal carriers but the CDW phase distortions almost neutralize the total charge. At other regimes, the lines of the zero CDW amplitude flash across the sample working as phase slips. The studies were inspired by, and can be applied to experiments on mesa-junctions in $NbSe_3$ and $TaS_3$ (Yu.I. Latyshev et al in proceedings of ECRYS 2008 and 2011).

## 1. Introduction: CDW at a junction

It is supposed that in conventional junction and tunneling devices the applied voltage doesn't modify the electronic states, just only shifting their positions and fillings of bare ones. But in correlated systems, particularly with a spontaneous symmetry breaking, the electronic spectra are formed self-consistently via electron-electron or electron-lattice interactions. These effects can modify the ground state, the spectra and even the very nature of carriers and collective modes, which are transformed following changes in concentration of electrons near junctions. As a result, charge storage and a current conduction become different entities rather than the same electrons. These effects came to a broad attention only very recently with the goal of controlled phase transformations at surfaces by the electrostatic doping requiring for extreme strengths of the electric field [1].

---

[*] Corresponding author. e-mail: kirova@lps.u-psud.fr

The CDWs are particularly attractive because here the reconstruction of the junction takes place at moderate experimentally attainable electric fields. The problem came to attention first in theory [2,3], then in experiment [4], and it became a must in view of decisive experimental demands [5,6] and in relation to other surface sensitive experiments [7,8]. The junction reconstruction in CDWs goes via appearance of topological defects (dislocations [9,10] as electronic vortices, as we shall call them below) with more microscopic solitons [11] as their cores.

We have already devoted studies and publications [12-14] to these problems. The numerical modeling was performed by the energy minimization for ground states under electrostatic voltage, by solutions of stationary PDEs for a system with running constant currents, solutions of time-dependent PDEs for transient processes recovering cascades of multiple vortices with a final stabilization for a few of them. While a kind of the Ginzburg-Landau (GL) or the time-dependent GL (TDGL) equations for the complex order parameter $\Psi$ of the CDW was in the core of the model, they were greatly complicated in several aspects: the higher nonlinearity of the TDGL equation itself; coupling of $\Psi$ with the normal carriers *n* which bring their own nonlinearity, retardations and dissipation via the diffusion equation; and particularly coupling to the electric potential $\Phi$ which brings the long range forces. This is the worst for the numerical work. Still, the simulations were always successful to an extent that they could be performed for realistic physical parameters and in actual sophisticated geometries of experiments.

Nevertheless, there is a serious demand for another, further complicated development which is described in this article and in [15]. In essence and by definition, the GL approach assumes integration over fermions (intrinsic carriers) participating in formation and distraction of the symmetry breaking, so that only $\Psi$ is left explicitly. Even in statics, the GL equation can be derived for a small gap which takes place only near the transition line. Moreover, the TDGL equations, whatever for the superconductivity or for the CDW [16], can be derived only for a dirty metal when the scattering by impurities suppresses the quasi-particle gap completely (while still leaving alive the order parameter amplitude). This is not the regime which we are interested in and what is demanded by the experiment.

The question is not just about precising some qualitatively apparent forms and results. The partitions of the collective and the normal components in densities of the charge and the current change qualitatively, and that is particularly pronounced near the cores of moving vortices.

## 2. Anomalous equations and their interpretation

The CDW deformation $\sim\Delta\cos(Qx+\varphi)$ is described by the complex order parameter $\Psi=A\exp(i\varphi)$, $A=\Delta/\Delta_0$ where $2\Delta_0$ is the CDW gap at T=0. Two types of normal carriers may coexist with the CDW: the intrinsic carries $n_{in}=(n_e,n_h)$ (as electrons and holes above and below the gap) and extrinsic ones $n_{ex}$.

Extrinsic carriers do not participate in the CDW formation and they are coupled with the CDW only via the Coulomb potential Φ; their potential energy is eΦ. These carriers belong to other electronic bands like pockets in NbSe$_3$ which example we shall imply. Their other sources can be non-gaped parts of an incompletely nested Fermi surface like in TbTe$_3$, etc. Intrinsic carriers exist in all realizations of, even if at low T they need to be activated across the gap. Their spectrum is formed by the CDW and their energies are displaced when the Fermi level E$_F$ breathes up and down with expansions/contractions of φ. Their total potential energy

$$e\Phi + \frac{\hbar v_F}{2}\frac{\partial \varphi}{\partial x}$$

adds the coupling with CDW phase deformations. (From now on, we include the electronic charge e into the potential Φ, all energies and temperature are measured in units of Δ$_0$, the length will be still in nm.)

*2.1. Equations*

We start with the following form of the local energy functional (see [15] on hints of derivation)

$$W = \frac{\hbar v_F}{4\pi}\left[\left(\frac{\partial A}{\partial x}\right)^2 + \beta^2\left(\frac{\partial A}{\partial y}\right)^2 + \left(\frac{\partial \varphi}{\partial x}\right)^2 + \beta^2 A^2 \left(\frac{\partial \varphi}{\partial y}\right)^2\right] +$$
$$\frac{e}{\pi}\Phi\frac{\partial \varphi}{\partial x} + e\Phi n_{ex} + \left(e\Phi + \frac{\hbar v_F}{2}\frac{\partial \varphi}{\partial x}\right)n_{in} - \frac{\varepsilon s}{8\pi}(\nabla\Phi)^2 \qquad (1)$$

We should add to that the local free energy of carriers $F_{ex}(n_{ex})+F_{in}(A,n_e,n_h)$. The dissipative evolution is described by eqs. generated from the functional (1):

$$\frac{\partial}{\partial x}\left(\frac{\partial \varphi}{\partial x} + \frac{2}{\xi_0}\Phi + \pi(n_e-n_h)\right) + \beta^2 \frac{\partial}{\partial y}\left(A^2 \frac{\partial \varphi}{\partial y}\right) = \gamma_\varphi A^2 \frac{\partial \varphi}{\partial t} \qquad (2)$$

$$-\frac{\partial^2 A}{\partial x^2} - \beta^2 \frac{\partial^2 A}{\partial y^2} + \beta^2 A \left(\frac{\partial \varphi}{\partial y}\right)^2 + \frac{\partial F}{\partial A} = -\gamma_A \frac{\partial A}{\partial t} \qquad (3)$$

Here $\xi_0=\Delta_0/\hbar v_F$ and $\gamma_{A,\varphi}$ are the damping coefficients. $\gamma_\varphi$ is related to the sliding CDW conductivity [12,13] which value fixes the time scale 10$^{-13}$ sec which will be the unit of our dimensionless time henceforth.

The Poisson equation for the electric potential is

$$\nabla^2\Phi = -\frac{\xi_0}{r_0^2}\left(\frac{\partial \varphi}{\partial x} + \pi(n_{in} + n_{ex})\right) \qquad (4)$$

where $r_0$ is the Thomas-Fermi screening length of the parent metal and $n_{in}=n_e-n_h$

The kinetics of normal carriers is taken in the quasi-equilibrium approximation.

$$\nabla(\sigma\nabla\mu) = \frac{e^2}{s}\frac{\partial n_{in}}{\partial t} \; ; \; \mu = \zeta + \Phi + \frac{\hbar v_F}{2}\frac{\partial \varphi}{\partial x} \qquad (5)$$

Here $\mu$ is the electrochemical potential $\zeta=\zeta(n,T)=\partial F/\partial n$ is the local chemical potential, $\sigma=(\sigma_x,\sigma_y)$ with $\sigma_i\sim(n_e+n_h)$ is the anisotropic conductivity tensor.

For boundary conditions, we assume that the normal CDW stress and the normal electric field are zero. The last arbitrary condition secures the total electro-neutrality and provides confinement of the electric potential within the sample which is convenient for calculations. The normal flow of the normal current exists only at two source/drain boundaries. As a whole, the boundary conditions have a form

$$\left(\frac{\partial \varphi}{\partial x} + \frac{2}{\xi_0}\Phi + \pi(n_e - n_h)\right)v_x + A^2 \frac{\partial \varphi}{\partial y} v_y = 0$$

$$\nabla A \cdot \vec{v} = 0 \quad \nabla \Phi \cdot \vec{v} = 0 \quad \vec{v} \cdot \nabla \mu = 0$$

(6)

Here $\vec{v}$ is the unit vector normal to the boundary.

The above eqs. contain thermodynamic parameters: $F$ and its derivatives. At a finite temperature T they can be calculated only numerically, so for the modeling we employed analytical interpolating formulas. For $F_{in}$ we used the BCS-Peierls form generalized to interpolate between small and large values of $\zeta$.

$$F(A,\zeta) = A^2\left(\log(A^2 + 7(\zeta^2 + T^2)) - 1\right)\frac{2\pi}{\xi_0^2}$$

The minimum of $F(A,\zeta)$ at $A\neq 0$ is erased (as it happens inside the vortex core) when $\zeta$ (hence $n_{in}$) is above a critical value. $n_e(\zeta,T) = n_h(-\zeta,T)$ were also given by formulas interpolating between large and small values of $|\zeta|$.

The dependence $\zeta(n)$ or $n(\zeta)$ defines dimensionless normal and collective particle densities:

$$\rho_n = \frac{1}{N_F}\frac{dn}{d\zeta} \quad \text{and} \quad \rho_c = 1 - \rho_n ; \quad N_F = \frac{2}{\pi\hbar v_F}$$

$N_F$ is is the density of states of parent metal at $E_F$. In the metallic phase by definition $\rho_n=1$, then $\rho_c=0$, approaching from the CDW phase as $\rho_c\sim\Delta^2$.

*2.2. Equations anomalies and their interpretation*

Notice that the terms with $\partial_x\varphi$, $\partial_{xx}\varphi$ and $(\partial_x\varphi)^2$ in eqs. (1), (2), (4) do not contain the usually supposed factor $A^2$. They are non analytic in the order parameter $\Psi$ and cannot be derived perturbatively; formally they appear because of the chiral anomaly [15].

Unlike the GL theory, all expressions containing $\partial_x\varphi$ are singular. Even the innocent eq. (3) for A is not normal: A couples conventionally with $\partial_y\varphi$ but there is no complementary coupling with $\partial_x\varphi$ because there was no cross-term in the energy (1).

But taking all eqs. in ensemble, we can notice, even if not explicitly, that the normal counterpart reacts negatively to variations of φ erasing the bare collective contribution in such a way that in terms with $\partial_x\varphi$ the factors *1* become $\rho_c=1-\rho_n$.

To illustrate how the anomalous eqs. (2-5) can yield the "normal" GL form, we can use the approximation of a local thermodynamic equilibrium μ=0. Suggest also that all deviations from equilibrium are small, hence the energy (1) is quadratic, and eqs. (2-5) are linear in all fields,

$\rho_c, \rho_n$ =cnst. Then we can eliminate the variable $n_{in}$ ("integrate out the fermions"). We arrive at a GL energy from which we show below only problematic terms with $\partial_x\varphi$ and $\Phi$:

$$\rho_c \frac{\hbar v_F}{4\pi}\left(\frac{\partial\varphi}{\partial x}\right)^2 + \frac{\rho_c}{\pi}\Phi\frac{\partial\varphi}{\partial x} - \frac{\rho_n}{\pi\hbar v_F}\Phi^2 - \frac{r_0^2}{\pi\hbar v_F}(\nabla\Phi)^2$$

The balance of the last two terms defines the expected length $l_{scr}=r_0/\sqrt{\rho_n}$ of screening by normal carriers. Particularly noticeable is the factor $\rho_c$ which appears in the first two terms as the reduction factors for elastic modulus and as the effective charge. With $\rho_c \sim A^2$ at small $A$, we recover in this limit the correct factors of the GL eqs. where the carriers have been excluded from the beginning. Then it is tempting, and looks quite intuitive, to use commonly accepted definitions of the charge and the current densities even in the nonlinear regime as

$$\pi n_c = \rho_c \partial_x \varphi; \quad \pi j_c = -\rho_c \partial_t \varphi \qquad (7)$$

But at a closer inspection, that violates the charge conservation law if all fields depend on both $x$ and $t$ which is well important for the motion of vortices:

$$\pi \frac{dn_c}{dt} = \frac{\partial \rho_c}{\partial x}\frac{\partial \varphi}{\partial t} - \frac{\partial \varphi}{\partial x}\frac{\partial \rho_c}{\partial t} \neq 0$$

In a nonlinear spacio-temporal regime, there is no explicit way to define the charge and current densities via the order parameter alone; we should use additively the bare collective and normal contributions as in the RHS of eq. (4). The collective charge $\partial_x\varphi/\pi$ does not depend either on the CDW amplitude $A$ or on population of carriers.

## 3. Numerical modelling.

We employed two different program environments: COMSOL and FreeFem++. We performed the numerical solution of above equations for two types of geometry (see Fig.1): rarely with the experimental one with slits and usually with the simple rectangular one. We modeled eqs. (2-6) in their full form, and also in the limit of the infinite conductivity $\sigma\rightarrow\infty$ for both components when eq. (5) is reduced to $\mu=0$.

Eqs. (2-5) were tried in two forms of the order parameters: i. $(A,\varphi)$-form with $\Psi= A\exp(i\varphi)$, and ii. (u,v)-form with $\Psi=(u+iv)$. The two forms are equivalent geometrically but not topologically. The $(A,\varphi)$ form leads to compact expressions and it runs very well. Its principle disadvantage is that it can yield the phase only as a multi-valued continuous function with no allowance for $2\pi$ jumps which are necessary to get an unbound vortex. That did not limit us in most of cases because the vortices usually stayed attached to the sample boundary. The (u,v)-form results in quite ugly eqs. which nevertheless were treatable by the programs. The resulting phase was defined periodically as $\tan(\varphi)$=v/u which accepts the $2\pi$ jumps, hence there is no constraints for detachment of vortices which was demonstrate earlier [12-14].

Different forms of eqs. and of choosing dependable variables and thermodynamic functions work differently sometimes, and the programs could crash at moments of vortex nucleation. But remarkably, the successful runs always give compatible results both in their shapes and critical parameters.

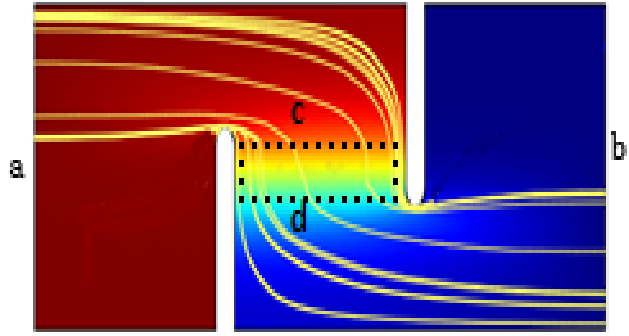

Figure 1. The real experimental mesa geometry with overlapping slits cut transversely to chains, and its active rectangular bridge indicated by dashed lines. Distributions of µ (the color density) and of the normal current (the stream lines) are shown for a sub-threshold voltage applied between sides *a* and *b*. In the simplified rectangular geometry the voltage is supposed to be applied to the sides c and d of the rectangular bridge.

## 3.1. Rectangular geometry, intrinsic carriers

For the rectangular geometry the CDW chains are oriented along the *x* direction, the voltage is applied across the CDW chain in *y* direction. For eqs. in both forms $(A,\varphi)$ and $(u,v)$ the results are mostly similar, Fig.2. In both cases the program was stable; it was run till the time $10^8$ when no more evolution could be seen.

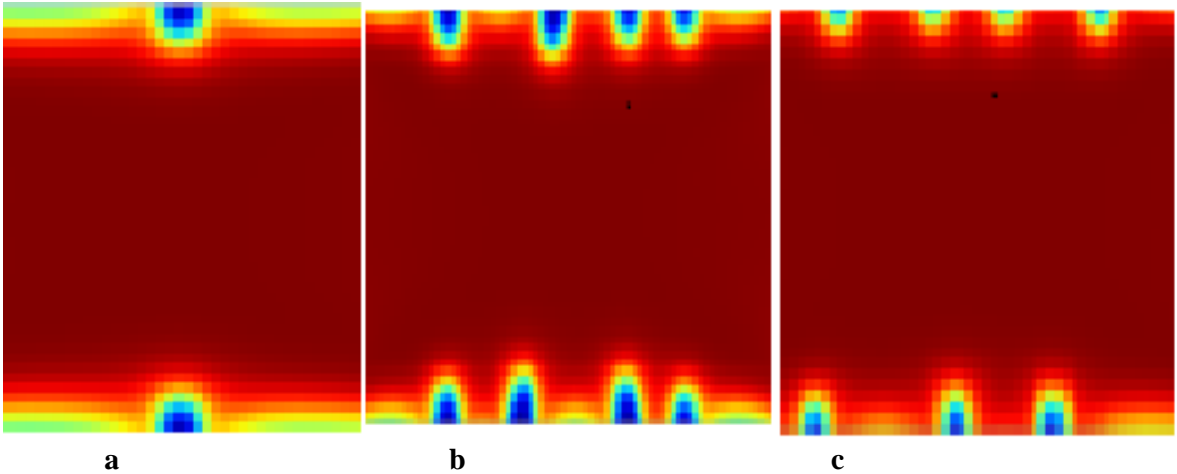

Fig.2. Creation of boundary vortices in the approximation µ=0. Colour density is given by A(x,y): from red for A=1 in the bulk to blue for A≈0 in the vortex core. **a**: $V=1.28\Delta_0$ - identical results for $(A,\varphi)$ and $(u,v)$ forms; **b**: $V=1.6\Delta_0$ in the $(A,\varphi)$ form; **c:** $V=1.6\Delta_0$ in the $(u,v)$ form.

A stable pair of vortices situated symmetrically at the upper and lower boundaries (Fig.2a) was found for V above nearly the same thresholds $V_{th1}=1.26\Delta_0$ for the $(A,\varphi)$ form and $V_{th2}=1.28\Delta_0$ for the $(u,v)$ form. With increasing V, more vortices appear above the higher thresholds. For the $(A,\varphi)$ form, four pairs appear at $V>V_{th2}=1.6\Delta_0$. For the $(u,v)$ form at $V=1.4\Delta_0$, four vortices appear at the upper boundary and only three ones at the lower boundary.

This non-symmetry of edges is not quite a program deficiency because the final number of vortices is a remnant of initial large number of nucleations, most of which disappear by the time $\sim 10^5$. While the $2\pi$ circulation of the phase could not be seen for the glued vortex, the phase jump across the vortex cores is observed for the path along the boundary: $\Delta\varphi=4.24=0.67*2\pi$.

A higher precision is necessary to understand if the vortex sits just at the surface or already is detached inwards the depth. That was clarified using another program environment: the FreeFem++. The program was sufficiently stable to solve the complete set of equations (2)-(5) for the bulky (u,v) scheme without using the approximation $\mu=0$.

We used the inter-chain coupling $\beta=0.1$, $\sigma_{\bar{x}}=\sigma_{CDW}$, $\sigma_x/\sigma_y=10^2$. For $V=1.3\Delta_0$ a pair of vortices was nucleated similar to Fig.2a. But with a further increase of V, the vortices detach from the surface inwards the bulk. The distance from the surface is small 0.8nm, but definitely final as it is seen in Fig. 3 and very clearly in Fig.4c. More vortices are nucleated with increasing voltage. Their maximum number we have been able to introduce in FreeFEM before divergence of the simulation was two pairs being created at $V=1.7\Delta_0$.

For creation of the first pair of vortices the critical voltage obtained by FreeFem is compatible with that one found by Comsol; but for the two vortex pairs the thresholds are different.

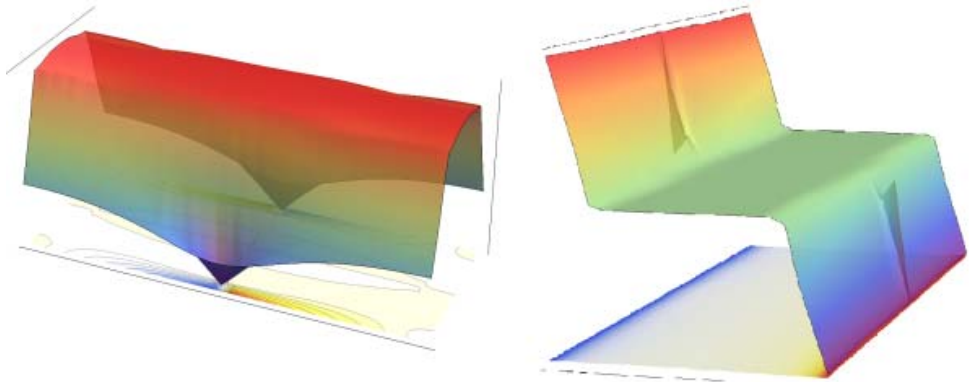

Figure 3 Left: 3D plot for A and contour plots for $\varphi$ after nucleation and stabilization of the pair of vortices in FreeFEM. Right:Density of intrinsic carriers with peaks at the cores.

The CDW amplitude, Fig.3, is almost constant in the bulk with a homogeneous depression along the boundaries within the layer of a width ~7nm which is the scale of $\xi_0$. On that background, A goes to zero approaching the vortex core. This result is in accordance with results obtained with Comsol, Fig. 4.

Peaks of n and $\zeta$ at the vortex cores stand out the background of zero which shows important variation of particle density at the vortex cores; the signs at the boundaries are opposite. Variations of the total charge density are also present but very small. The density of intrinsic carriers, see Figs. 3b and 4a,b is negligible in the bulk, having a substantial homogeneous increase at the surface layer, in accordance with the potential drop there. It further increases sharply in the region of the vortex core.

The basic reason for that is the necessity to maintain approximately the electro-neutrality by compensating the growth of *q* towards the core. In its turn, growing *n* or equivalently $\zeta$ suppress the energy minimum p
osition towards *A=0*.

This information from the 3D plots is clarified at linear plots over the sample cross-section passing through the vortex cores, Fig. 4. The comparison of boundary values at points x far away (left panel) and at the vortex core (middle panel) are the following. A=0.85 and 0.3, µ=0.64 $\zeta$= -0.5 and -1.5. Φ=0.64 everywhere as fixed by the boundary condition. Notice three times increase in value of $\zeta$, hence in concentration of normal particles. The total charge is almost compensated. For the detached vortex (right panel) A passes truly through zero but the potentials grow towards the nearby boundary and are not very sensitive to the point where A=0. Both $\zeta$ and Φ grow in magnitude but with a different sign, leaving µ depressed but flat in the area of depressed A. Remind that the FreeFEM computations where done beyond the approximation µ=0 so the the boundary condition was imposed upon µ, not φ.

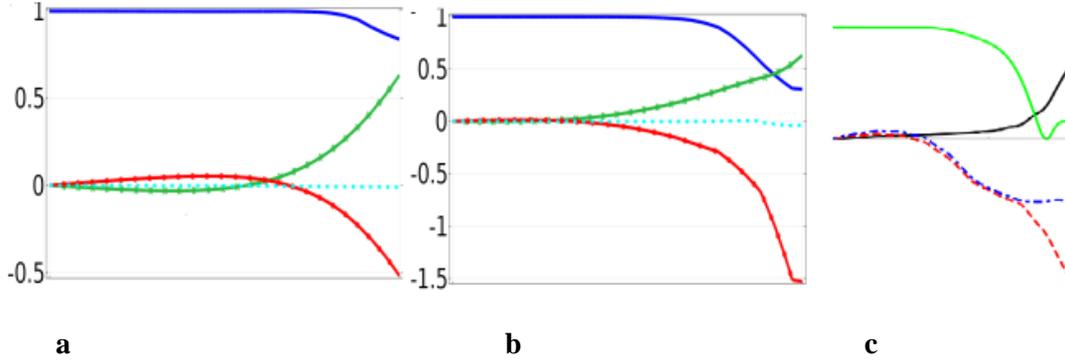

a  b  c

Figure 4. Plots along cross-sections *x=cnst* for cases of Figs. 2 and 3. a: *x=80*, far away from the vortex core; b: (Comsol) and c: (FreeFem) across the vortex core x=0. a and b: Φ (green,dotted), $\zeta$ (red), A (blue), the total charge (dashed cyan). c: Φ (black), $\zeta$ (red dashed), A (green), µ (blue dash-dotted).

### 3.2. Flashing phase slips, extrinsic carriers

Here we briefly report preliminary observations of a truly dynamical regime of flashing phase slips. It was obtained in the above given approximation µ=0 and only for the (A,φ) form. Moreover, the extrinsic carriers were needed to be taken into account, probably to provide a more efficient screening.

Fig. 5 gives a snapshot at t=12589 for the color density plot of the amplitude (left panel). We see several types of flashes (from the left to the right): almost complete, branching, complete, retreating ones. Inside each trace, *A* is suppressed truly to zero, Φ drops to -3 (its mean bulk value is -0.5), $\zeta$ grows strongly but differently among the flashes: from 3 to 7 ($\zeta \approx 0$ in the bulk).

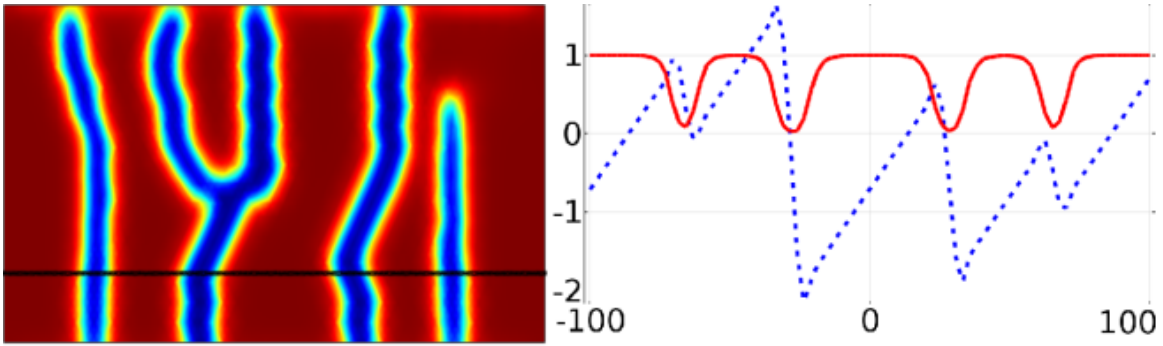

Figure 5. *Left:* density plots of A. *Right:* x dependences along the cross-section marked by the black line in the left panel. Solid plot – A, dotted plot - φ.

At the right panel of Fig.5, we see the deeps of the amplitude passing by which the phase promptly drops to resume it linear growth in between (keeping the same gradient).

*3.3.    Slits geometry: preliminary results.*

In Fig.6 we presented a solution for the real slit geometry, see Fig.1. We solved the full equation system in the u-v method without approximations. Only the intrinsic carriers were taken into account. The program crashed after the vortices have been nucleated but results obtained by that moment are reliable. The two-vortex configuration see Fig 6, left panel, was computed for $\sigma_x=10\sigma_{CDW}$, $\sigma_y=0.01\sigma_{CDW}$, $\beta=1$, at $V=0.8\Delta_0$. The dips of A, probably the future vortices, are created in the outer vicinity of tips of each slit. That may be related to the high current density in these areas as is evident in Fig.1a. These preliminary hints show that the processes in the slit geometry may be more complicated than in the rectangular idealization.

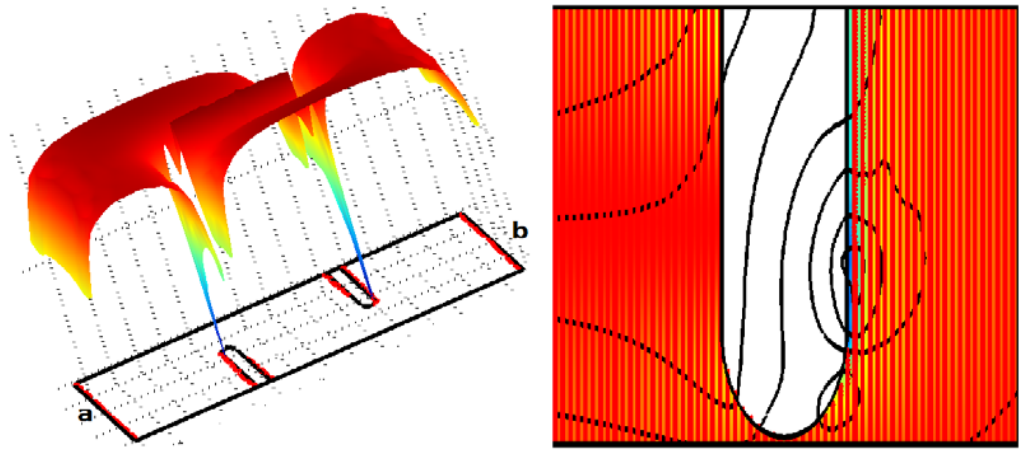

Figure 6 *Left*: At the threshold voltage for nucleation of vortices. 3D plot with colors for A and contour plots for the phase. *Right:* amplified vicinity of the tip of the right slit. The deep of A is seen as the blue region in the color density). Black lines are equipotentials Φ=cnst.

## 4. Conclusions

We have studied the reconstruction of the CDW state under the applied transverse voltage in the internal junction. The calculations were performed for parameters close to experiments on NbSe$_3$.

We employed new equations derived in [15] and explained in this article. The new approach was called by problems of definition and conservation of collective and normal charge densities. That required for keeping tracks of normal carriers instead of prescription of a GL-kind theory to integrate them out at the early stage. The price was a non-analytical structure of equations which brings challenges to numerical procedures.

Nevertheless, time-evolution equations were solved numerically for a restricted geometry in two spatial dimensions. The simulations were performed by the finite element method implemented via programs COMSOL and FreeFem++. We have obtained creation and subsequent multiplications of electronic vortices (the CDW dislocations) at boundary layers of junctions. In special cases the dynamic phase slips were seen as lines of zero CDW amplitude flashing rapidly across the junction.
.

**References**

[1]     Proceedings of IMPACT12: "Electronic States and Phases Induced by Electric or Optical Impacts", S. Brazovskii and N.Kirova Eds., Eur. Phys. J. Special Topics **222** (2013).
[2]     S. Brazovskii and S. Matveenko, Sov. Phys. JETP **74** (1992) 864.
[3]     N.Kirova, and S.Brazovskii, J. Physique IV **12** (2002) 173.
[4]     T. L. Adelman, S. V. Zaitsev-Zotov, and R. E. Thorne, Phys. Rev. Lett. **74** (1995) 5264.
[5]     Y. I. Latyshev, et al, Phys. Rev. Lett. **95** (2005) 266402.
[6]     Y. I. Latychev, et al, Phys. Rev. Lett. **96** (2006) 116402.
[7]     D. Le Bolloc'h, et al, Phys. Rev. Lett. **100** (2008) 096403.
[8]     E. Pinsolle, et al, Phys. Rev. Lett. **109** (2012) 256402.
[9]     D. Feinberg and J. Friedel, J. de Phys. **49** (1988) 485.
[10]    S. Brazovskii and S. Matveenko Sov. Phys. JETP **72** (1991) 860.
[11]    S. Brazovskii, et al, Phys. Rev. Lett. **108** (2012) 096801; Ch.Brun in this volume.
[12]    T. Yi, et al, J. Supercond. Nov. Mag. **25** (2012) 1323.
[13]    T. Yi, et al, Physica B **407** (2012) 1839.
[14]    T. Yi, N. Kirova, S. Brazovskii, Eur. Phys. J. Special Topics **222** (2013) 1035.
[15]    T. Yi, et al, J. Supercond. Nov. Magn. **1411** (2015) 1868 arXiv: 1411.1868.
[16]    L. P. Gor'kov, JETP Lett. **38** (1983) 87.